\documentclass[conference]{IEEEtran}
\IEEEoverridecommandlockouts
\usepackage{cite}
\usepackage{amsmath,amssymb,amsfonts}
\usepackage{algorithmic}
\usepackage{graphicx}
\usepackage{textcomp}
\usepackage{xcolor}
\usepackage{booktabs} 
\usepackage{caption}
\usepackage{lipsum} 
\usepackage{booktabs}

\usepackage{hyperref} 
\def\BibTeX{{\rm B\kern-.05em{\sc i\kern-.025em b}\kern-.08em
    T\kern-.1667em\lower.7ex\hbox{E}\kern-.125emX}}
    
\begin{document}

\title{Federated Quantum Kernel-Based Long Short-term \\ Memory for Human Activity Recognition

\thanks{\IEEEauthorrefmark{9} Corresponding Author: 
ycchen1989@ieee.org, kuoenjui@nycu.edu.tw. The views expressed in this article are those of the authors and do not represent the views of Wells Fargo. This article is for informational purposes only. Nothing contained in this article should be construed as investment advice. Wells Fargo makes no express or implied warranties and expressly disclaims all legal, tax, and accounting implications related to this article.}
}

\author{
\IEEEauthorblockN{
    Yu-Chao Hsu\IEEEauthorrefmark{2}\IEEEauthorrefmark{3},
    Jiun-Cheng Jiang\IEEEauthorrefmark{2}\IEEEauthorrefmark{4},
    Chun-Hua Lin\IEEEauthorrefmark{4}\IEEEauthorrefmark{2},
    Wei-Ting Chen\IEEEauthorrefmark{5},
    Kuo-Chung Peng\IEEEauthorrefmark{4}\IEEEauthorrefmark{2},\\
    Prayag Tiwari\IEEEauthorrefmark{7},
    Samuel Yen-Chi Chen\IEEEauthorrefmark{6}\IEEEauthorrefmark{9},
    En-Jui Kuo\IEEEauthorrefmark{8}\IEEEauthorrefmark{9}
}
\IEEEauthorblockA{\IEEEauthorrefmark{2} National Center for High-Performance Computing, NARlabs, Hsinchu, Taiwan}
\IEEEauthorblockA{\IEEEauthorrefmark{3} Cross College Elite Program, National Cheng Kung University, Tainan, Taiwan}
\IEEEauthorblockA{\IEEEauthorrefmark{4} Department of Physics and Center for Theoretical Physics, National Taiwan University, Taipei, Taiwan}
\IEEEauthorblockA{\IEEEauthorrefmark{5} Department of Computer Science \& Engineering, University of California San Diego, USA}
\IEEEauthorblockA{\IEEEauthorrefmark{7} School of Information Technology, Halmstad University, Sweden}
\IEEEauthorblockA{\IEEEauthorrefmark{6} Wells Fargo, New York, NY, USA}
\IEEEauthorblockA{\IEEEauthorrefmark{8} Department of Electrophysics, National Yang Ming Chiao Tung University, Hsinchu, Taiwan}
}


\maketitle

\begin{abstract}
In this work, we introduce the Federated Quantum Kernel-Based Long Short-term Memory (Fed-QK-LSTM) framework, integrating the quantum kernel methods and Long Short-term Memory into federated learning.
Within Fed-QK-LSTM framework, we enhance human activity recognition (HAR) in privacy-sensitive environments and leverage quantum computing for distributed learning systems.
The DeepConv-QK-LSTM architecture on each client node employs convolutional layers for efficient local pattern capture, this design enables the use of a shallow QK-LSTM to model long-range relationships within the HAR data.
The quantum kernel method enables the model to capture complex non-linear relationships in multivariate time-series data with fewer trainable parameters.
Experimental results on RealWorld HAR dataset demonstrate that Fed-QK-LSTM framework achieves competitive accuracy across different client settings and local training rounds.
We showcase the potential of Fed-QK-LSTM framework for robust and privacy-preserving human activity recognition in real-world applications, especially in edge computing environments and on scarce quantum devices.
\end{abstract}


\section{Introduction}

In recent years, \textit{privacy-preserving computing} has become an active research field, especially when the data is in related to human-beings \cite{ryoo2016privacypreservinghumanactivityrecognition,10285538,sym16010089}.
Driven by this motivation, Human Activity Recognition (HAR) is a classification problem we want to solve with, while a deep learning architecture, DeepConvLSTM, proposed by Ordonez et al.~\cite{s16010115} has demonstrated the extraordinary performance for HAR in previous researches \cite{Bock_2021,zhou2025}.

Moreover, to enable privacy-preserving computing on machine learning (ML) model, one familiar method is federated learning (FL).
Federated learning has emerged as a decentralized machine learning paradigm that enables multiple clients to collaboratively train a shared global model without exposing their raw data \cite{mcmahan2023communicationefficientlearningdeepnetworks, kairouz2021advancesopenproblemsfederated}.
This approach addresses critical privacy and data sovereignty concerns in sensitive domains such as healthcare, finance, and mobile applications.
In FL, clients perform local training using their own datasets and only communicate model updates with a central server or aggregation mechanism, thus mitigating data leakage risks and reducing communication overhead.
Recent advances in FL have demonstrated its efficacy in diverse applications, especially where data is distributed and heterogeneous across edge devices \cite{li2020convergencefedavgnoniiddata, Nishio_2019, https://doi.org/10.48550/arxiv.1806.00582}.

Quantum computing (QC) emerges as a novel computational approach with potential of superior advantages over classical counterpart in some tasks, specifically, machine learning, quantum chemistry and combinatorial optimization problems \cite{Biamonte_2017, Abbas_2021, PhysRevResearch.5.043216, liu2022hybridgatebasedannealingquantum}.
In particular, quantum machine learning (QML) draws many attentions recently from its success in various applications by leveraging quantum superposition and entanglement to embed classical data more efficiently\cite{Schuld_2021, liu2024quantumtrainrethinkinghybridquantumclassical, hsu2025quantum, DEVADAS2025103318,chen2022quantum,hsu2025qae}.
Furthermore, to integrate FL with QML, leading to quantum federated learning (QFL), opens new directions for decentralized, privacy-preserving intelligent systems \cite{chen2021federatedquantummachinelearning, liu2024federatedquantumtrainbatchedparameter,liu2025federated}.

In this work, we are going to demonstrate how to implement DeepConvLSTM with quantum kernel-based long short-term memory (QK-LSTM) in Section~\ref{sec:lstm}.
In Section~\ref{sec:qfl}, we further introduce DeepConv-QK-LSTM within quantum federated learning framework, which allows the model to learn HAR in a privacy-preserving manner.
In Section~\ref{sec:result}, we showcase our numerical results and validate the feasibility and capability of our approach.
Finally, we conclude the paper in Section~\ref{sec:dis}.

\begin{figure*}[!t]
    \centering
    \vspace{-12pt}
    \includegraphics[width=0.8\textwidth]{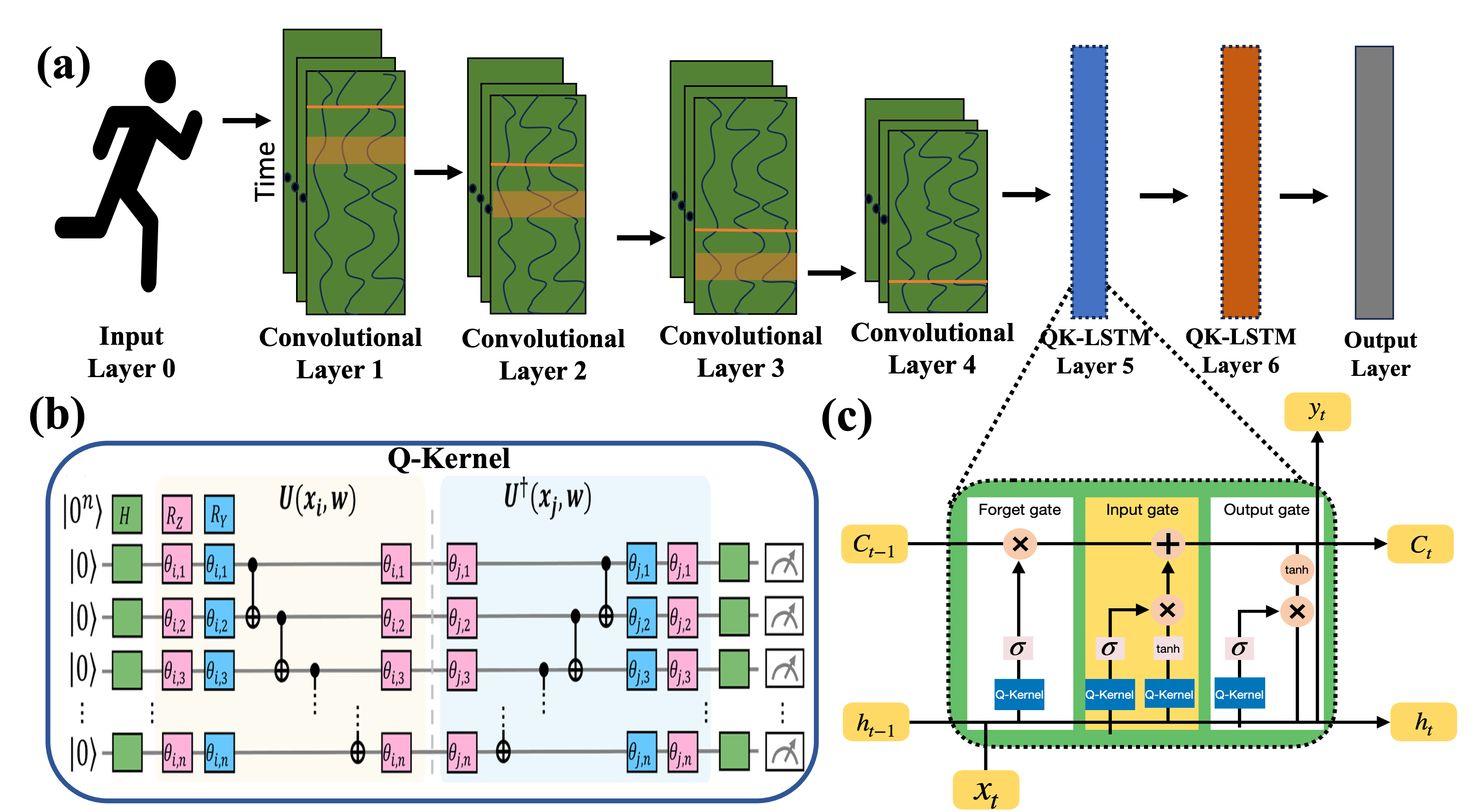}
    \vspace{-6pt}
    \caption{
    \textbf {Illustration of the DeepConv-QK-LSTM model.} 
    (a) The input human activity sensor data is first processed through a series of convolutional layers to extract local temporal features. 
    (b) The extracted features are encoded into quantum states using angle encoding, and quantum kernel functions are evaluated via an embedding circuit \( U(x_i, w) \) and its adjoint \( U^\dagger(x_j, w) \), forming the basis for kernel-based similarity computation. 
    (c) These kernel values are used within the QK-LSTM unit to compute gate activations (forget, input, and output gates), enabling quantum-enhanced modeling of temporal dependencies.
    }
    \label{DeepConv-QK-LSTM}
    \vspace{-12pt}
\end{figure*}

\section{Long Short-Term Memory}\label{sec:lstm}
\subsection{Classical Long Short-Term Memory}

Long Short-Term Memory (LSTM) has significantly advanced the fields of machine learning and sequential data modeling\cite{hochreiter1997long}. By integrating input, forget and output gates within a unified memory cell, LSTM networks offer a stable mechanism for learning long-term relationships in sequential data. LSTM networks have found extensive applications across multiple fields, including time series forecasting\cite{kilinc2024multimodal,siami2019performance,cao2019financial,wen2023time}, natural language processing\cite{chen2016enhanced,wang2015learning,yao2018improved}, and chemical applications \cite{ertl2017silico,xu2023novel}, demonstrating strong capabilities in tasks ranging from language translation and sensor prediction to molecular property estimation.
In this study, we utilize LSTM networks for Human Activity Recognition, due to their strong capability in modeling temporal dependencies within multivariate sensor signals\cite{bock2021improving}. The sequential nature of HAR data, often collected from accelerometers\cite{kwapisz2011activity,lara2013survey}, gyroscopes\cite{anguita2013public}, and other wearable sensors\cite{lara2013survey}, requires models that can effectively capture both short-term motion patterns and long-term contextual dependencies.

\subsection{Quantum Kernel-Based Long Short-Term Memory}
In our study, to improve the performance and reduce the model parameters of LSTM networks in HAR task, we introduce the QK-LSTM model \cite{hsuLSTM2025quantum,hsu2025quantum}, a novel hybrid machine learning architecture that integrates the strengths of classical LSTM networks with quantum kernel method \cite{thanasilp2024exponential,paine2023quantum,chen2024validating}.
By integrating quantum kernel methods into the LSTM, QK-LSTM takes advantage of the expressive power of high-dimensional feature spaces of quantum computing to significantly improve the representational capacity of the model.
Formal mathematical formulation of QK-LSTM Cell is given by:



\begin{subequations}
\allowdisplaybreaks
\vspace{-3pt}
\begin{align}
f_t &= \sigma\left( \sum_{j=1}^{N} \beta_j^{(f)} \, \kappa^{(f)}(v_t, v_j) \right), \\
i_t &= \sigma\left( \sum_{j=1}^{N} \beta_j^{(i)} \, \kappa^{(i)}(v_t, v_j) \right), \\
\hat{C}_t &= \tanh\left( \sum_{j=1}^{N} \beta_j^{(C)} \, \kappa^{(C)}(v_t, v_j) \right), \\
C_t &= f_t \odot C_{t-1} + i_t \odot \hat{C}_t, \\
o_t &= \sigma\left( \sum_{j=1}^{N} \beta_j^{(o)} \, \kappa^{(o)}(v_t, v_j) \right), \\
h_t &= o_t \odot \tanh\left( C_t \right),
\end{align}
\vspace{-3pt}
\end{subequations}
\noindent
where:
\begin{itemize}
    \item  \(v_t = [h_{t-1}; x_t] \in \mathbb{R}^{n + m}\) represents the concatenation of input \(x_t\) at time-step \(t\) and the hidden state \(h_{t-1}\) from the previous time-step\(t-1\), $\odot$ denotes the element-wise (Hadamard) product.
    \item \( \beta_j^{(f)} \), \( \beta_j^{(i)} \), \( \beta_j^{(C)} \), and \( \beta_j^{(o)} \) are trainable coefficients corresponding to each gate's quantum kernel,
    \item \( \kappa^{(f)}(\cdot, \cdot) \), \( \kappa^{(i)}(\cdot, \cdot) \), \( \kappa^{(C)}(\cdot, \cdot) \), and \( \kappa^{(o)}(\cdot, \cdot) \) denote the quantum kernel functions tailored to the respective gates.
    
\end{itemize}

\subsection{DeepConv-QK-LSTM}
Some studies have shown that DeepConvLSTM architectures are particularly effective for HAR tasks and that shallow LSTM layers are often sufficient when preceded by deep convolutional feature extractors\cite{s16010115,bock2021improving,zhou2025,nooruddin2023design}. Motivated by these findings, we employ a multilayer Deep Convolutional (DeepConv) later to extract informative features from the Human Activity Recognition (HAR) dataset~\cite{sztyler2016body}. The DeepConv network is designed to capture local temporal patterns and reduce dimensionality before passing the extracted feature representations to the QK-LSTM layer.

We denote the input time-series HAR data signal as a multivariate sequence
\(X = [x_1, x_2, \dots, x_T] \in \mathbb{R}^{T \times d}\), where \( T \) is the number of time steps and \( d \) denotes the number of input channels. At the \( l \)-th convolutional layer, 1D temporal filters are applied to extract the local time features. The activation of the \( k \)-th output channel at time \( t \) is computed by:
\begin{equation}
h_t^{(l, k)} = ReLU \left( \sum_{i=0}^{K-1} \sum_{c=1}^{C^{(l-1)}} w_{i, c}^{(l,k)} \cdot h_{t+i}^{(l-1, c)} + b^{(l,k)} \right),\vspace{-12pt}
\end{equation}
\vspace{-1pt}
where:
\begin{itemize}
    \item \( h_t^{(l, k)} \) denotes the output of the \( k \)-th filter at time \( t \) in layer \( l \),
        \item \( K \) is the temporal kernel size,
    \item \( C^{(l-1)} \) is the number of channels in the previous layer,
    \item \( w_{i, c}^{(l,k)} \) and \( b^{(l,k)} \) are learnable weights and biases, respectively,
    \item \( ReLU(\cdot) \) is a nonlinear activation function,
    \item The initial input is defined as \( h_t^{(0)} = x_t \).
\end{itemize}
\vspace{-1pt}

After passing through the number of \( L \) convolutional layers, the final extracted representation is \(\phi(X) = \left[\phi(x_1), \dots, \phi(x_T)\right] \in \mathbb{R}^{T \times p}\), where \( \phi(x_t) \in \mathbb{R}^p \) denotes the learned feature vector at time step \( t \), and \( p \) is the number of output channels. These feature vectors serve as input to the QK-LSTM layer to enable temporal modeling across multiple scales. At each time step \( t \), the QK-LSTM receives a concatenated input vector \( v_t = [h_{t-1}; \phi(x_t)] \in \mathbb{R}^{n + p} \), which combines the previous hidden state \( h_{t-1} \) with the current convolutional feature \( \phi(x_t) \), thereby incorporating both historical context and deep local features into the quantum kernel computation for QK-LSTM. The overall architecture is illustrated in Fig.~\ref{DeepConv-QK-LSTM}.

\section{Federated Learning}\label{sec:qfl}

\begin{figure}[t]
    \centering
    \vspace{-12pt}
    \includegraphics[width=0.5\textwidth]{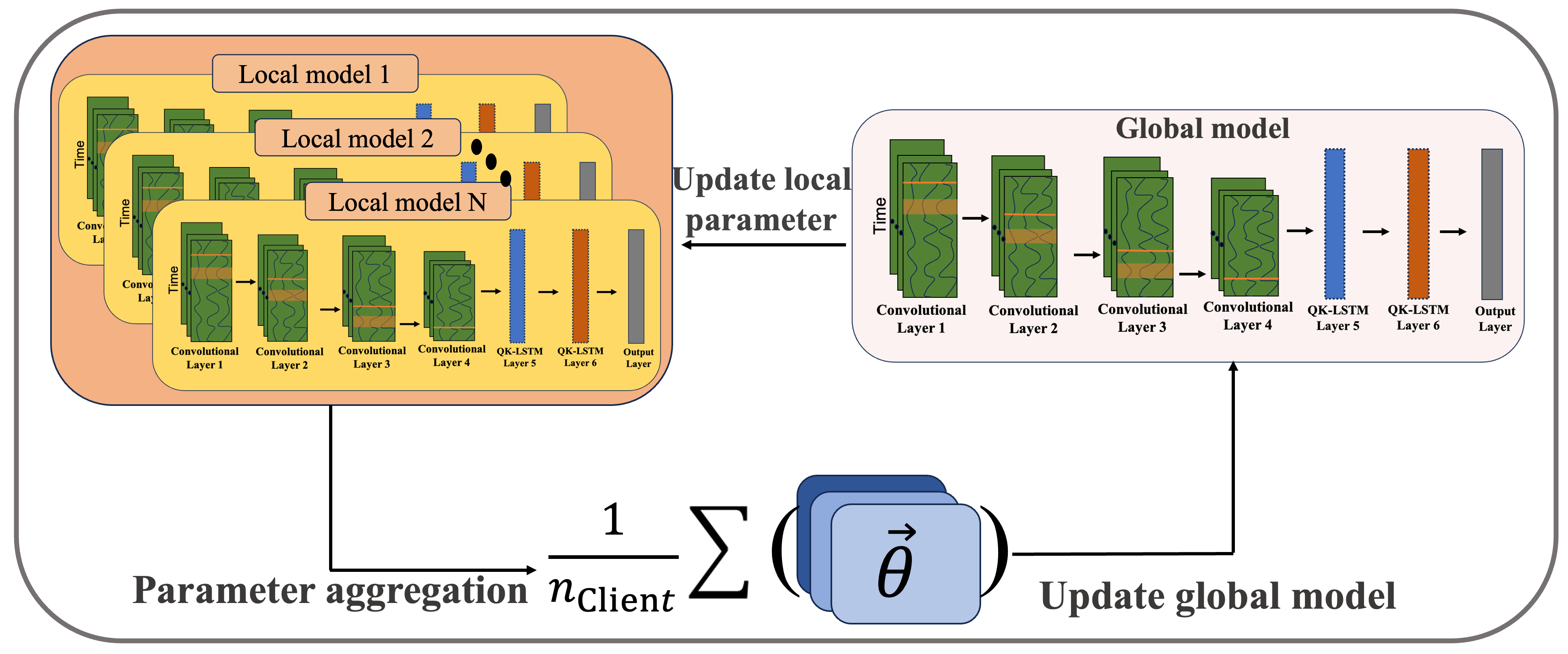}
    \vspace{-16pt}
    \caption{
        \textbf{Schematic representation of the Fed-QK-LSTM.}
        }
    \label{Fed-QK-LSTM}
    \vspace{-14pt}
\end{figure}

\subsection{Federated Learning}
Federated learning has undergone rapid development since its initial formulation by Google in 2016\cite{McMahan2016FederatedLO} as a communication-efficient solution for decentralized deep learning under privacy constraints.
Rather than uploading raw data to a central server, FL frameworks enable collaborative training through an iterative process where each client locally optimizes model parameters and transmits only model updates for global aggregation.
One of the most widely adopted methods in this setting is the Federated Averaging (FedAvg) algorithm, which synchronizes local models by averaging their weights at a central aggregator across communication rounds.
This approach has demonstrated robustness under data heterogeneity and limited communication bandwidth \cite{mcmahan2023communicationefficientlearningdeepnetworks, kairouz2021advancesopenproblemsfederated}. 

\subsection{Federated QK-LSTM}\label{FQKLSTM}
With the advent of quantum computing and quantum machine learning, researchers have started to explore federated learning in the quantum context which is quantum federated learning. In simple terms, QFL adapts the principles of FL to scenarios where either the data, the model, or the computing resources are quantum in nature.
In this work, we propose a hybrid quantum-classical sequence modeling framework, namely Federated Quantum Kernel-based Long Short-term Memory (Fed-QK-LSTM).
This framework integrates the DeepConv-QK-LSTM model as the primary client architecture within a federated learning environment, thereby enabling quantum-enhanced sequence modeling capabilities.
Each client node encodes local convolutional features \( \phi(x_t) \in \mathbb{R}^p \) into a quantum state \( \vert \psi(\theta)\rangle \) via angle encoding, and computes quantum kernel values for optimizing parameters of the local target model.
During each training round, clients perform local model updates on the DeepConv-QK-LSTM model parameters using their private local datasets.
After local training, the obtained parameters will be transmitted to the central aggregator for global aggregation, as depicted in the Fig.~\ref{Fed-QK-LSTM}

{In Federated Quantum Kernel‑Based LSTM (Fed‑QK‑LSTM), each client \(k\) trains both (a) the classical convolutional and LSTM weights and (b) the parameters associated with the quantum‑kernel which include the kernel coefficients \( \beta_j^{(f,k)} \), \( \beta_j^{(i,k)} \), \( \beta_j^{(C,k)} \), \( \beta_j^{(o,k)} \). Each client uses simulated quantum circuits to locally compute quantum kernel values over \(n_k\) data samples. These kernels act as fixed, nonlinear feature maps that project input vectors into a higher-dimensional Hilbert space, enabling similarity measurement in this transformed space. Importantly, model's trainable parameters remain real-valued and are used to form linear combinations of the kernel outputs. Consequently, model updates can be aggregated using standard techniques from classical federated learning, such as Federated Averaging (FedAvg). }

This process enables collaborative learning across distributed devices while preserving user privacy, as each client locally trains the DeepConv-QK-LSTM model on its own data.
By exchanging only model parameters with the central server, communication overhead is reduced and sensitive information remains protected.
Integration of quantum kernel methods enhances the global model's ability to capture complex temporal patterns in multi-sensor time series data.
Compared to former QFL frameworks, our approach requires fewer parameters by compact representation of the quantum kernel.

\section{Numerical Result and Discussion}\label{sec:result}
\begin{figure}[t]
    \centering
    \vspace{-12pt}
    \includegraphics[width=0.48\textwidth]{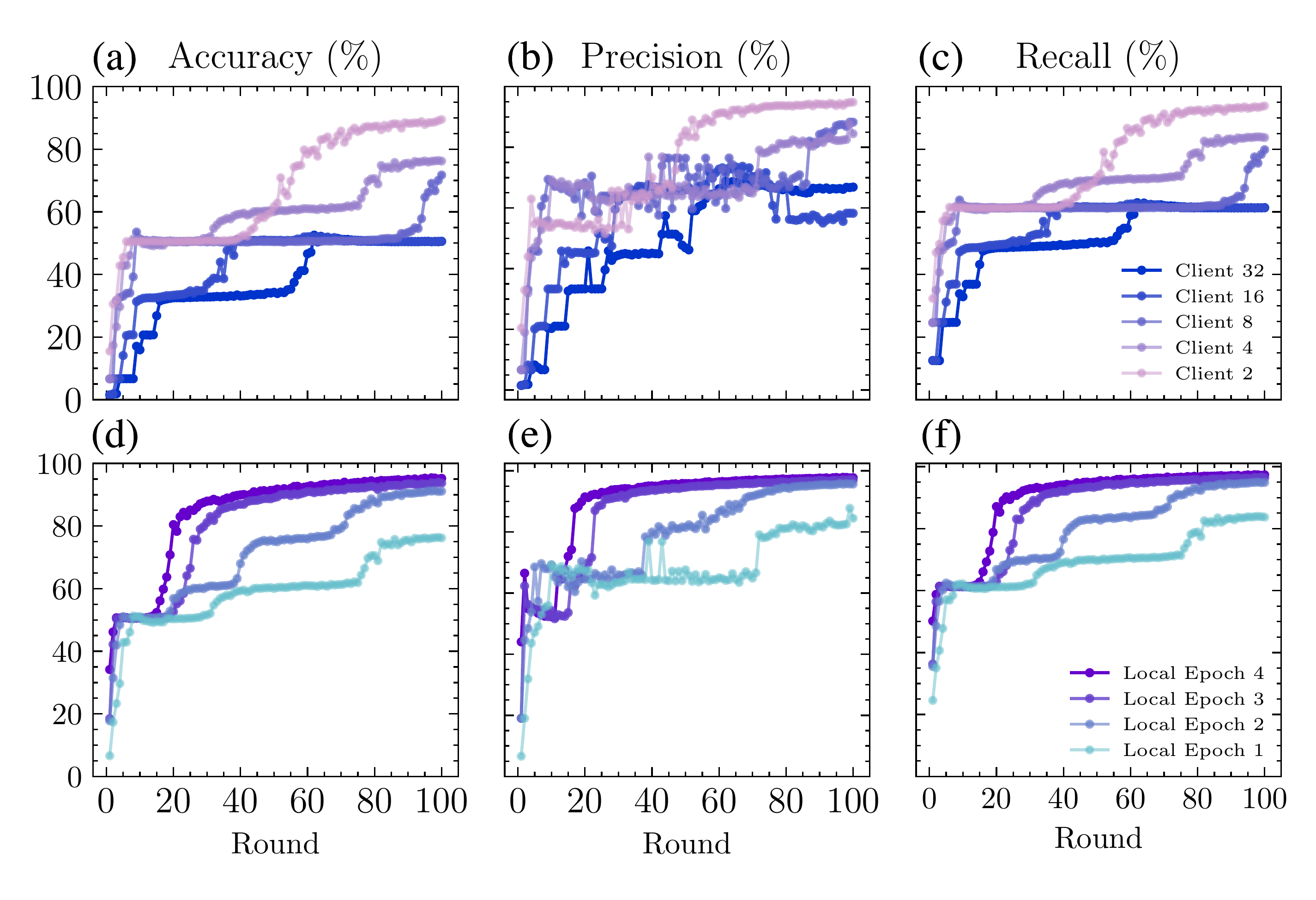}
    \vspace{-12pt}
    \caption{
    \textbf{Experimental results of federated learning with different number of clients and different number of local epochs of each client.}
    The first row indicates the experiment on different numbers of clients involved in the federated learning, and each local epoch is one.
    The second row is our model training on a fixed number of 3 clients and with different numbers of local epochs.
    }
    \label{fig:result}
    \vspace{-14pt}
\end{figure}

We construct the quantum kernel based on Block-Product State (BPS) wavefunctions\cite{suzuki2024quantum,chen2024validating} and numerically simulate it with PennyLane \cite{bergholm2022pennylaneautomaticdifferentiationhybrid} and PyTorch \cite{paszke2019pytorchimperativestylehighperformance}.
The configuration of our model has 2 layers of QK-LSTM each with 64 units and convolutional layer have 64 filters with filter width 11.
We use Adam \cite{kingma2017adammethodstochasticoptimization} for optimization, with hyperparameters as follows: \textit{learning rate} = $10^{-4}$, \textit{weight decay} = $10^{-4}$, \textit{dropout probability} = 0.5, and a fixed random seed.
We assess the effectiveness of Fed-QK-LSTM with a popular HAR datasets, RealWorld HAR (RWHAR) \cite{sztyler2016body}, where the sampling rate is set as 50 Hz.
The model train with different number of clients $\in{\{2, 4, 8, 16, 32\}}$ and train on different number of local epoch $\in{\{1, 2, 3 ,4\}}$.
The results are reported in Fig.~\ref{fig:result}.


To examine the applicability of the Fed-QK-LSTM framework, we tested it with the HAR dataset.
As Fig.~\ref{fig:result} (a), (b), and (c) show, all three evaluation criteria demonstrate a high level of performance.
When the client number increases, the curve shows a diminishing slope, indicating a slowing rate of change.
This rapid initial increase indicates that with fewer clients, the model can quickly adapt and improve, likely due to each client holding more data or contributing more significant updates during each round of aggregation \cite{cowlishaw2025balancing}.
This can be addressed by running more global rounds and increasing dataset size.
Furthermore, increasing the local epoch does speed up the learning process as well.
As shown in the second row of Fig.~\ref{fig:result} (d), (e), and (f), the behavior of the three diagrams exhibits a trend similar to that observed in the previous experiment.
However, it is more demanding of computational resources on the local devices.

\begin{table}[!b]
\vspace{-12pt}
\centering
\caption{\small \textbf{Comparative Performance of Fed-LSTM and Fed-QK-LSTM Models.}}
\label{tab:compar}
\small
\setlength{\tabcolsep}{8pt}
 \centering
    \begin{tabular}{@{}lccc@{}}
\toprule
\textbf{Metric}     & \textbf{Fed-LSTM} & \textbf{Fed-QK-LSTM (ours)} \\
\midrule
Accuracy   & 0.90      & \textbf{0.95}    \\
Precision  & 0.95      & \textbf{0.97}    \\
Recall     & 0.94      & \textbf{0.97}    \\
F1-score   & 0.94      & \textbf{0.97}    \\
\midrule
\textbf{Trainable Parameters} & 202,696& \textbf{137,996}  \\
\bottomrule
\end{tabular}
\end{table}

{To compare the performance between the quantum and classical models, we conducted experiments using 3 clients and 4 local training epochs.
Experimental results across Fed-LSTM and Fed-QK-LSTM are
summarized in Table~\ref{tab:compar}.
Fed-QK-LSTM consistently outperforms Fed-LSTM in classification accuracy on HAR dataset.
In addition to higher accuracy, Fed-QK-LSTM also achieves superior performance in terms of precision, recall, and F1-score, while requiring fewer trainable parameters.}


\section{Conclusion}\label{sec:dis}
In this work, we introduce an application of the QK-LSTM in federated learning using high-sensitivity datasets, such as HAR datasets.
Each client node employs QK-LSTM and a local model to generate local parameters updated based on the local datasets and aggregated at the central aggregator.

Our experiments with Fed-QK-LSTM on HAR datasets demonstrate the feasibility of next-generation quantum-enhanced machine learning methods in real-world applications. We vary the number of clients to examine the relationship between client population size and performance metrics such as accuracy, precision, and recall. Additionally, we assess model performance under different numbers of local epochs.

The experimental results reveal that as the client population grows, the performance curves for these three metrics tend to flatten, reflecting decreased learning efficiency. Moreover, increasing local epochs serves as an effective method to enhance learning efficiency. As shown in Table ~\ref{tab:compar}, our approach achieves superior results compared to the classical method.

We demonstrate the practicality of employing quantum machine learning approaches for processing privacy-sensitive data.
Our results confirm the feasibility of applying quantum-augmented federated learning in real-world scenarios, and pave the way for operational deployment in scalable quantum computing applications and distributed learning environments.



\section*{Acknowledgment}
The authors would like to thank the National Center for High-performance Computing of Taiwan for providing computational and storage resources. YCH thanks Dr. Douhan Justin Yeh for his discussion.

\bibliographystyle{ieeetr}
\bibliography{reference} 

\begin{thebibliography}{10}

\bibitem{ryoo2016privacypreservinghumanactivityrecognition}
M.~S. Ryoo {\em et~al.}, ``Privacy-preserving human activity recognition from
  extreme low resolution,'' 2016.

\bibitem{10285538}
A.~Jain {\em et~al.}, ``Privacy-preserving human activity recognition system
  for assisted living environments,'' {\em IEEE Transactions on Artificial
  Intelligence}, vol.~5, no.~5, pp.~2342--2357, 2024.

\bibitem{sym16010089}
L.~Wang {\em et~al.}, ``A blockchain-based privacy-preserving healthcare data
  sharing scheme for incremental updates,'' {\em Symmetry}, vol.~16, no.~1,
  2024.

\bibitem{s16010115}
F.~J. Ordóñez {\em et~al.}, ``Deep convolutional and lstm recurrent neural
  networks for multimodal wearable activity recognition,'' {\em Sensors},
  vol.~16, no.~1, 2016.

\bibitem{Bock_2021}
M.~Bock {\em et~al.}, ``Improving deep learning for har with shallow lstms,''
  in {\em 2021 International Symposium on Wearable Computers}, UbiComp ’21,
  ACM, Sept. 2021.

\bibitem{zhou2025}
H.~Zhou {\em et~al.}, ``Efficient human activity recognition on edge devices
  using deepconv lstm architectures,'' {\em Scientific Reports}, vol.~15, 04
  2025.

\bibitem{mcmahan2023communicationefficientlearningdeepnetworks}
H.~B. McMahan {\em et~al.}, ``Communication-efficient learning of deep networks
  from decentralized data,'' 2023.

\bibitem{kairouz2021advancesopenproblemsfederated}
P.~Kairouz {\em et~al.}, ``Advances and open problems in federated learning,''
  2021.

\bibitem{li2020convergencefedavgnoniiddata}
X.~Li {\em et~al.}, ``On the convergence of fedavg on non-iid data,'' 2020.

\bibitem{Nishio_2019}
T.~Nishio and R.~Yonetani, ``Client selection for federated learning with
  heterogeneous resources in mobile edge,'' in {\em ICC 2019 - 2019 IEEE
  International Conference on Communications (ICC)}, IEEE, May 2019.

\bibitem{https://doi.org/10.48550/arxiv.1806.00582}
Y.~Zhao {\em et~al.}, ``Federated learning with non-iid data,'' 2018.

\bibitem{Biamonte_2017}
J.~Biamonte {\em et~al.}, ``Quantum machine learning,'' {\em Nature}, vol.~549,
  p.~195–202, Sept. 2017.

\bibitem{Abbas_2021}
A.~Abbas {\em et~al.}, ``The power of quantum neural networks,'' {\em Nature
  Computational Science}, vol.~1, p.~403–409, June 2021.

\bibitem{PhysRevResearch.5.043216}
S.-K. Chou {\em et~al.}, ``Accurate harmonic vibrational frequencies for
  diatomic molecules via quantum computing,'' {\em Phys. Rev. Res.}, vol.~5,
  p.~043216, Dec 2023.

\bibitem{liu2022hybridgatebasedannealingquantum}
C.-Y. Liu and H.-S. Goan, ``Hybrid gate-based and annealing quantum computing
  for large-size ising problems,'' 2022.

\bibitem{Schuld_2021}
M.~Schuld, R.~Sweke, and J.~J. Meyer, ``Effect of data encoding on the
  expressive power of variational quantum-machine-learning models,'' {\em
  Physical Review A}, vol.~103, Mar. 2021.

\bibitem{liu2024quantumtrainrethinkinghybridquantumclassical}
C.-Y. Liu {\em et~al.}, ``Quantum-train: Rethinking hybrid quantum-classical
  machine learning in the model compression perspective,'' 2024.

\bibitem{hsu2025quantum}
Y.-C. Hsu {\em et~al.}, ``Quantum kernel-based long short-term memory for
  climate time-series forecasting,'' in {\em 2025 International Conference on
  Quantum Communications, Networking, and Computing (QCNC)}, pp.~421--426,
  IEEE, 2025.

\bibitem{DEVADAS2025103318}
R.~M. Devadas and S.~T, ``Quantum machine learning: A comprehensive review of
  integrating ai with quantum computing for computational advancements,'' {\em
  MethodsX}, vol.~14, p.~103318, 2025.

\bibitem{chen2022quantum}
S.~Y.-C. Chen {\em et~al.}, ``Quantum convolutional neural networks for high
  energy physics data analysis,'' {\em Physical Review Research}, vol.~4,
  no.~1, p.~013231, 2022.

\bibitem{hsu2025qae}
Y.-C. Hsu, K.-C. Chen, T.-Y. Li, and N.-Y. Chen, ``Quantum adaptive excitation
  network with variational quantum circuits for channel attention,'' {\em arXiv
  preprint arXiv:2507.11217}, 2025.

\bibitem{chen2021federatedquantummachinelearning}
S.~Y.-C. Chen and S.~Yoo, ``Federated quantum machine learning,'' 2021.

\bibitem{liu2024federatedquantumtrainbatchedparameter}
C.-Y. Liu and S.~Y.-C. Chen, ``Federated quantum-train with batched parameter
  generation,'' 2024.

\bibitem{liu2025federated}
C.-Y. Liu {\em et~al.}, ``Federated quantum-train long short-term memory for
  gravitational wave signal,'' {\em arXiv preprint arXiv:2503.16049}, 2025.

\bibitem{hochreiter1997long}
S.~Hochreiter and J.~Schmidhuber, ``Long short-term memory,'' {\em Neural
  computation}, vol.~9, no.~8, pp.~1735--1780, 1997.

\bibitem{kilinc2024multimodal}
H.~C. Kilinc {\em et~al.}, ``Multimodal fusion of optimized gru--lstm with
  self-attention layer for hydrological time series forecasting,'' {\em Water
  Resources Management}, vol.~38, no.~15, pp.~6045--6062, 2024.

\bibitem{siami2019performance}
S.~Siami-Namini {\em et~al.}, ``The performance of lstm and bilstm in
  forecasting time series,'' in {\em 2019 IEEE International conference on big
  data (Big Data)}, pp.~3285--3292, IEEE, 2019.

\bibitem{cao2019financial}
J.~Cao, Z.~Li, and J.~Li, ``Financial time series forecasting model based on
  ceemdan and lstm,'' {\em Physica A: Statistical mechanics and its
  applications}, vol.~519, pp.~127--139, 2019.

\bibitem{wen2023time}
X.~Wen and W.~Li, ``Time series prediction based on lstm-attention-lstm
  model,'' {\em IEEE access}, vol.~11, pp.~48322--48331, 2023.

\bibitem{chen2016enhanced}
Q.~Chen {\em et~al.}, ``Enhanced lstm for natural language inference,'' {\em
  arXiv preprint arXiv:1609.06038}, 2016.

\bibitem{wang2015learning}
S.~Wang and J.~Jiang, ``Learning natural language inference with lstm,'' {\em
  arXiv preprint arXiv:1512.08849}, 2015.

\bibitem{yao2018improved}
L.~Yao and Y.~Guan, ``An improved lstm structure for natural language
  processing,'' in {\em 2018 IEEE international conference of safety produce
  informatization (IICSPI)}, pp.~565--569, IEEE, 2018.

\bibitem{ertl2017silico}
P.~Ertl {\em et~al.}, ``In silico generation of novel, drug-like chemical
  matter using the lstm neural network,'' {\em arXiv preprint
  arXiv:1712.07449}, 2017.

\bibitem{xu2023novel}
B.~Xu {\em et~al.}, ``A novel second-order learning algorithm based
  attention-lstm model for dynamic chemical process modeling,'' {\em Applied
  Intelligence}, vol.~53, no.~2, pp.~1619--1639, 2023.

\bibitem{bock2021improving}
M.~Bock {\em et~al.}, ``Improving deep learning for har with shallow lstms,''
  in {\em Proceedings of the 2021 ACM International Symposium on Wearable
  Computers}, pp.~7--12, 2021.

\bibitem{kwapisz2011activity}
J.~R. Kwapisz {\em et~al.}, ``Activity recognition using cell phone
  accelerometers,'' {\em ACM SIGKDD Explorations Newsletter}, vol.~12, no.~2,
  pp.~74--82, 2011.

\bibitem{lara2013survey}
O.~D. Lara and M.~A. Labrador, ``A survey on human activity recognition using
  wearable sensors,'' {\em IEEE Communications Surveys \& Tutorials}, vol.~15,
  no.~3, pp.~1192--1209, 2013.

\bibitem{anguita2013public}
D.~Anguita {\em et~al.}, ``A public domain dataset for human activity
  recognition using smartphones,'' in {\em Proceedings of the 21st European
  Symposium on Artificial Neural Networks (ESANN)}, 2013.

\bibitem{hsuLSTM2025quantum}
Y.-C. Hsu {\em et~al.}, ``Quantum kernel-based long short-term memory,'' in
  {\em 2025 IEEE International Conference on Acoustics, Speech, and Signal
  Processing Workshops (ICASSPW)}, pp.~1--5, IEEE, 2025.

\bibitem{thanasilp2024exponential}
S.~Thanasilp {\em et~al.}, ``Exponential concentration in quantum kernel
  methods,'' {\em Nature communications}, vol.~15, no.~1, p.~5200, 2024.

\bibitem{paine2023quantum}
A.~E. Paine, V.~E. Elfving, and O.~Kyriienko, ``Quantum kernel methods for
  solving regression problems and differential equations,'' {\em Physical
  Review A}, vol.~107, no.~3, p.~032428, 2023.

\bibitem{chen2024validating}
K.-C. Chen {\em et~al.}, ``Validating large-scale quantum machine learning:
  Efficient simulation of quantum support vector machines using tensor
  networks,'' {\em Machine Learning: Science and Technology}, 2024.

\bibitem{nooruddin2023design}
S.~Nooruddin, ``On the design of efficient deep learning methods for human
  activity recognition in resource constrained devices,'' Master's thesis,
  University of Waterloo, 2023.

\bibitem{sztyler2016body}
T.~Sztyler and H.~Stuckenschmidt, ``On-body localization of wearable devices:
  An investigation of position-aware activity recognition,'' in {\em 2016 IEEE
  international conference on pervasive computing and communications (PerCom)},
  pp.~1--9, IEEE, 2016.

\bibitem{McMahan2016FederatedLO}
H.~B. McMahan {\em et~al.}, ``Federated learning of deep networks using model
  averaging,'' {\em ArXiv}, vol.~abs/1602.05629, 2016.

\bibitem{suzuki2024quantum}
T.~Suzuki {\em et~al.}, ``Quantum support vector machines for classification
  and regression on a trapped-ion quantum computer,'' {\em Quantum Machine
  Intelligence}, vol.~6, no.~1, p.~31, 2024.

\bibitem{bergholm2022pennylaneautomaticdifferentiationhybrid}
V.~Bergholm {\em et~al.}, ``Pennylane: Automatic differentiation of hybrid
  quantum-classical computations,'' 2022.

\bibitem{paszke2019pytorchimperativestylehighperformance}
A.~Paszke {\em et~al.}, ``Pytorch: An imperative style, high-performance deep
  learning library,'' 2019.

\bibitem{kingma2017adammethodstochasticoptimization}
D.~P. Kingma {\em et~al.}, ``Adam: A method for stochastic optimization,''
  2017.

\bibitem{cowlishaw2025balancing}
R.~Cowlishaw {\em et~al.}, ``Balancing centralisation and decentralisation in
  federated learning for earth observation-based agricultural predictions,''
  {\em Scientific Reports}, vol.~15, p.~10454, 2025.

\end{thebibliography}
\end{document}